\documentclass[manuscript,screen]{acmart}
\RequirePackage{hyperref}  % Ensure hyperref is loaded first
\usepackage{hyperxmp}      % Load hyperxmp immediately after hyperref

\usepackage{multirow}
\usepackage{lscape}
\usepackage{booktabs}
\setcopyright{acmcopyright}
\copyrightyear{2018}
\acmYear{2018}
\acmDOI{XXXXXXX.XXXXXXX}

\acmConference[Conference acronym 'XX]{Make sure to enter the correct
  conference title from your rights confirmation emai}{June 03--05,
  2018}{Woodstock, NY}
\acmBooktitle{Woodstock '18: ACM Symposium on Neural Gaze Detection,
 June 03--05, 2018, Woodstock, NY} 
\acmPrice{15.00}
\acmISBN{978-1-4503-XXXX-X/18/06}

\begin{document}

\title{Conversational Agents for Older Adults' Health: A Systematic Literature Review}

\author{Jiaxin An}
\email{jiaxin.an@utexas.edu}
\orcid{0000-0003-2793-6469}
\affiliation{%
  \institution{University of Texas at Austin}
  \city{Austin}
  \state{Texas}
  \country{USA}
  \postcode{78712}
}
\author{Siqi Yi}
\email{sukie_yi@utexas.edu}
\affiliation{%
  \institution{University of Texas at Austin}
  \city{Austin}
  \state{Texas}
  \country{USA}
  \postcode{78712}
}

\author{Yao Lyu}
\email{yaolyu@umich.edu}
\affiliation{%
  \institution{University of Michigan}
  \city{Ann Arbor}
  \state{Michigan}
  \country{USA}
  \postcode{48103}
}

\author{Houjiang Liu}
\email{liu.ho@utexas.edu}
\affiliation{%
  \institution{University of Texas at Austin}
  \city{Austin}
  \state{Texas}
  \country{USA}
  \postcode{78712}
}

\author{Yan Zhang}
\email{yanz@utexas.edu}
\affiliation{%
  \institution{University of Texas at Austin}
  \city{Austin}
  \state{Texas}
  \country{USA}
  \postcode{78712}
}

%% \renewcommand{\shortauthors}{An, et al.}

\begin{abstract}
There has been vast literature that studies Conversational Agents (CAs) in facilitating older adults' health. The vast and diverse studies warrants a comprehensive review that concludes the main findings and proposes research directions for future studies, while few literature review did it from human-computer interaction (HCI) perspective. In this study, we present a survey of existing studies on CAs for older adults' health. Through a systematic review of 72 papers, this work reviewed previously studied older adults' characteristics and analyzed participants' experiences and expectations of CAs for health. We found that (1) Past research has an increasing interest on chatbots and voice assistants and applied CA as multiple roles in older adults' health. (2) Older adults mainly showed low acceptance CAs for health due to various reasons, such as unstable effects, harm to independence, and privacy concerns. (3) Older adults expect CAs to be able to support multiple functions, to communicate using natural language, to be personalized, and to allow users full control. We also discuss the implications based on the findings. 

%Recent advancements in generative AI have highlighted the potential of Conversational Agents (CAs) to improve older adults' health. Although previous studies have explored various applications, the complex challenges and opportunities of implementing CAs in senior healthcare remain under-examined. In this study, we present a systematic survey

% this work reviewed previously studied older adults' demographic characteristics. It also  
\end{abstract}



\maketitle

\section{Introduction} \label{sec:introduction}

By 2050, older adults are projected to make up 16.0\% of the global population, placing enormous pressure on healthcare systems to accommodate the needs of an aging population \citep{united_nations_department_of_economic_and_social_affairs_population_division_world_2020}. To address this growing challenge, conversational agents (CAs) --- also referred to as chatbots, conversational AI, dialogue systems, or virtual assistants \cite{maharjan_hear_2019, garg_last_2022, car_conversational_2020, laranjo_conversational_2018} --- are increasingly viewed as a promising technological intervention. CAs engage users in conversational interactions, primarily through text and spoken natural language, to assist various aspects of older adults' lives. Applications of CAs for older adults range from those support their daily activities, such as meal preparation \cite{czaja_designing_2019}, time management and home maintenance \cite{even_benefits_2022, czaja_designing_2019, even_benefits_2022}, to those provide healthcare support, including medication, preventative care, and disease management \cite{czaja_designing_2019, even_benefits_2022}. Additionally, the recent advancement of generative AI technologies, especially large language models (LLMs), presents promising opportunities to engage with older adults with more nuanced and contextually relevant responses compared to traditional CAs, such as rule-based or other predefined technical approaches. 

%Recent studies have applied LLMs to offer personalized health managements, such as facilitating health-related communications between healthcare providers and patients \citep{Yang2024-fe} and motivating physical activity behavior change \citep{Jorke2024-qq}. By leveraging LLMs, healthcare systems can offer scalable, accessible, and responsive care, addressing older adults' needs more effectively \citep{Mahmood2023-yi, Ma2023-qe}. 

While increasing evidence has demonstrate the positive impact of this technology on older adults' physical health and mental well-being, challenges remain in their use of CAs. This includes the difficulty in learning to use \cite{even_benefits_2022, zhang_role_2024}, interaction breakdowns\cite{even_benefits_2022}, privacy concerns, and lack of trust \cite{even_benefits_2022, zhang_role_2024}. More importantly, due to the sensitive nature of medical data and the potential high stakes involved in health-related decisions, the complexities and challenges of applying CAs for older adults' health are significant. A systematic review from Human-Computer Interaction (HCI) perspective is needed to thoroughly examine the research landscape of CAs for older adults' health, including the emerging diverse forms and modalities of these technologies and various healthcare scenarios where CAs could be applicable.

% Note that technology literacy refers to the level of competence and familiarity individuals have with technology, typically developed through long-term exposure and continuous learning. However, older adults often struggle to keep pace with significant advancements in new technologies due to factors such as cognitive decline, limited education, or a lack of ongoing updates to their skills.

In this paper, we seek to provide a systematic review of the CAs designed for older adults' health. This review will examine the current applications of CAs in supporting older adults' health and explore older adults' experiences and expectations of using CAs. By doing so, we aim to identify effective strategies for integrating CAs into healthcare practices and highlight areas where improvements are needed. In particular, we aimed to answer the following research questions:

% Research has investigated the potential of CAs in facilitating older adults' healthcare and reduce the burden of the healthcare system\cite{parmar_health-focused_2022,kocaballi_conversational_2020}. The studies applied CA for multiple purposes, such as self-management, therapy, health information collection, health monitoring, training and education, conseling and social engagement\cite{zhang_role_2024}. A growing number of studies found the positive effect of CA usage on older adults’ physical health and social wellbeing \cite{sin_avoiding_2022}. While challenges existed in older adults' interaction with CAs, including the 

% Existing literature reviews tried to identify the purposes, benefits and challenges of CA application either for healthy older adults \cite{even_benefits_2022} or focusing on the chatbot \cite{zhang_role_2024}, which results in the neglect of older adults with health conditions (e.g., dementia) and other types of CAs (e.g., embodied conversational agents (ECAs)).

% Additionally, there is no clear mapping or synthesis of themes for the effect of CAs on older adults' health, older adults' perception and their expectation of CAs for health. Our work aims to fill in this gap by providing a comprehensive systematic review of the CAs for older adults' health. In particular, we aimed to answer the following questions:

\begin{itemize}
    \item RQ1: \textbf{What are the CAs in older adults' health?}
    \item RQ2: \textbf{What are older adults' experiences of using CAs for health?}
    \item RQ3: \textbf{What are older adults' expectations of CA for health?}
\end{itemize}

%- Older adults' acceptance level of using CAs
%- Older adults' evaluation of CA
%- Concern of CA

% RQ3: What are older adults' expectation of CAs for their health? 
%- Expectation/need

% RQ4: What are the effects of CA on older adults' health?
%- N of studies investigate the outcome of CA on older adults' health
%- Type of outcomes: positive

%%RQ1: What are existing CAs (both prototypes and products) for older adults’ healthcare and their characteristics, such as roles, interactions, underpinning theories, and architecture?

%%RQ2: What contexts have been explored in research on CAs for older adults’ health, including temporal, spatial, user/personal, and situational contexts?

%%RQ3: What study designs have been used in existing research?

%%RQ4: Benefits and challenges of?

Our review identified 72 relevant articles published before 2024. In this review, we found that the studies shifted their focus from ECAs (Embodied Conversational Agents, one type of CA with human-like avatars) to voice assistants and chatbots. They applied CAs in multiple health domains as different roles, such as coach, companion, general health assistant, therapist, health information provider, and monitor. Our analysis further identified that the majority or studies were conducted in the Global North and recruited older adults who were females and highly-educated. With respect to older adults' experiences and expectations of CAs, we found a mixed effect of CAs on older adults' health; our findings highlight older adults' low acceptance of CAs and their concerns of using CAs: privacy, dependence, cost, stigmatization, usability, and information quality; we summarized older adults expectations of CAs as functionalities, natural conversations, personalization and sense of control. Regarding future research, we discussed four aspects of challenges in designing CAs for older adults' health: the heterogeneity of older adults, older adults' need of independence, their expectation of powerful CAs and the unclear effect of CAs. 

%To our knowledge, this is the first systematic literature review of older adults' use of CAs in health...

%This paper is structured as follows. In Section \ref{sec:related work}, we briefly review the how CAs are designed and used for health and the diverse CA applications designed for older adults. We then synthesize related survey that also focused on CAs for older adults' health and highlight the unique perspective of our study. In Section \ref{sec:methods}, we describe the systematic review process, including the literature search, criteria for study identification, and the qualitative coding methodology. In Section \ref{sec:findings}, we report the main findings, each addressing the research questions listed above. Subsequently, in Section \ref{sec:discussion}, we elaborate on the evolving research and propose new directions for future studies. Finally, we conclude in Section \ref{sec:conclusion}. 

Our contributions to the HCI literature are three-fold: First, we present a systematic survey of research on conversational agents for older adults, including specifying eligibility criteria, conducting survey, and screening irrelevant studies; second, we analyze the demographic information of older adult populations and their experiences and expectations of CAs; finally, we discuss the findings within the broader context of HCI for health and provide implications for future work.

\section{Related work} \label{sec:related work}

\subsection{Conversational Agents (CAs) for Health}
The first conversational agents (CAs) in health was ELIZA, which was developed in 1966 as a psychotherapist \citep{car_conversational_2020}. ELIZA is a text-based computer program that can identify keywords in users' responses and transform them based on rules set by developers \citep{weizenbaum_elizacomputer_1966}. \citet{car_conversational_2020} named such a rule-based CA as "simple CA" because the user is "restricted to predetermined options when answering questions posted by conversational agents, and there is little or no opportunity for free responses." Recent advancements in Artificial Intelligence (AI) have led to the development of various AI techniques, including traditional Machine Learning (ML), Natural Language Processing (NLP), and Computer Vision (CV), as well as hybrid models that integrate voice, text, and visual elements, and large language models (LLMs). These innovations have broadened the forms of new CAs. text-based chatbots use text-generation models to deliver free-form responses \citep{oh_empathy_2017}. Voice assistants employ text-to-speech (TTS) models to facilitate natural language conversations \citep{maharjan_hear_2019}. Virtual characters, such as embodied conversational agents (ECAs), utilize multimodal approaches to offer both verbal and visual interactions \citep{sebastian_changing_2017, zhou_virtual_2022}. 

Given this technological advancement, CAs have been employed to perform a variety of health-related tasks for diverse user groups. For example, regarding medical professionals, CAs can be used in training medical students \citep{montenegro_survey_2019}. For patients, CAs can provide healthcare information for patients (e.g., training, coaching or counseling, prevention, diagnosis) in different health domains (e.g., nutrition, therapy, and cardiology). Additionally, the collaborative use of CAs can improve existing healthcare practices by engaging a wider range of health-related user groups \citep{montenegro_survey_2019}. For example, CAs can address sexual and reproductive issues among adolescents \citep{rahman_adolescentbot_2021} and support team-based group therapy for mental health \citep{dosovitsky_bonding_2021, maharjan_hear_2019, narynov_chatbots_2021}.

While the benefits of using CAs in healthcare are evident, particularly in alleviating human resource shortages and offering user-friendly support for individuals with difficulties in communications or low eHealth literacy \citep{bickmore_using_2009}, technical limitations and ethical issues were also highlighted. For example, \citet{park_impact_2022} report that only a few CA applications in the market that adopt the new AI techniques mentioned above are fully compliant with health information privacy regulations and standards. Additionally, most CAs designed for health might not be fully capable of supporting emergency conditions, such as real-time crisis management \citep{Xiao2025-xx}, urgent medical advice or immediate intervention for life-threatening conditions \citep{Stieglitz2022-mv}. On the other hand, they might generate improper responses due to biased training data or undefined conversational flow. More importantly. the black-box nature of the algorithms imposes difficulties for the systems to be transparent and able to explain their outputs. As recent studies have found that CAs were more likely to be trusted by people who do not know how they work, this limitation amplifies potential negative impacts \citep{denecke_artificial_2021, miner_smartphone-based_2016}. This limitation can exacerbate potential negative effects, creating a double burden for older adults, who often face challenges in keeping up with rapid technological advancements. This is the research subject that we now turn.

\subsection{Designing CAs for Older Adults}
There has been significant growth in both academic research and industrial efforts focusing on designing CAs for older adults. Studies in this area have identified several factors that impact how older adults use these technologies. As various consumer products are available in the market (e.g., Amazon Echo\footnote{\url{https://www.amazon.com/Echo}} or Apple HomePod\footnote{\url{https://www.apple.com/homepod/}}), scholars began to investigate older adults' different attitudes and reactions towards using these products, finding that older adults generally recognize the benefits of using CAs to support independent living, such as setting reminders or providing hands-free access to information, particularly for those with disabilities \citep{upadhyay_studying_2023, kowalski_older_2019, pradhan_more-than-human_2022, trajkova_alexa_2020}. 
%Additionally, compared to younger generations, older adults seem to be more attentive to the nonverbal emotional behaviors, such as sad face, dead down or nodding \citep{hosseinpanah_empathy_2018}. This might be due to seniors' lack of experience in using CAs, thus, observing the agent showing emotional responses make them feel less patronized. As a result, they are more likely to perceive CAs as trustworthy. 

However, the challenges of adopting CAs for older adults vary widely, not only in terms of their direct limitations in capabilities but also due to other social factors. For instance, a year-long study conducted by \citet{upadhyay_studying_2023} observed that while CAs like Amazon Alexa can assist older adults with daily tasks, they fail to provide dynamic and ongoing conversations. Additionally, \citet{pradhan_use_2020} reported that older adults often struggle to remember or locate specific voice commands, which can limit their effective use of this technology. \citet{trajkova_alexa_2020} also noted that family members of older adults often have conflicting opinions on CAs; some view them as helpful tools for fostering independence, while others find them cumbersome, requiring negotiation with their family members. Moreover, older adults raised concerns about privacy protection for potential excessive dependence on using CAs \citep{trajkova_alexa_2020}.

While previous research have depicted the opportunities of CAs designed for health, as well as the benefits and challenges of adopting CAs in assisting older adults' daily activities, there remains a limited understanding of how CAs can be effectively applied to meet particular older adults' health needs. Health-related aging presents unique challenges compared to other age groups, including the complexities related to ethical compliance and health regulations specifically for older adults \citep{Mahoney2007-tq, Hellstrom2007-pb} and their increased physical and mental vulnerabilities \citep{Neves2015-pl, Suhaimi2022-jp, Pang2021-km} affecting the design of CAs. Additionally, as previous studies suggest \citep{trajkova_alexa_2020, pradhan_use_2020, upadhyay_studying_2023, kowalski_older_2019}, older adults' experiences and expectations of CAs are crucial for their adoption and effective use. The elderly may exhibit varying levels of trust, comfort, and familiarity with technology, which can significantly influence their willingness to adopt CAs, particularly in healthcare settings. In summary, the sensitive nature of medical data, the high stakes of health-related decisions, and the unique challenges older adults face when using CAs necessitate a more systematic understanding of the various human factors involved for the design of CAs.

\subsection{Related Survey of CAs for Older Adults' Health in HCI}
Although a few medical reviews assess the clinical applications of conversational agents (CAs), including their use for older adults \citep{Milne-Ives2020-pf, Bin-Sawad2022-ct}, there remains a significant gap in systematic surveys or reviews within the HCI community focusing specifically on CAs for older adults’ health. Medical reviews primarily emphasize clinical efficacy and health outcomes, whereas HCI reviews tend to focus on design implications related to user experience, accessibility, and other human factors.

Specifically, only two literature reviews have focused on the role of CAs in the context of older adults’ health. For instance, \citet{even_benefits_2022} conducted a scoping review of 23 papers, synthesizing the benefits and challenges of commercial CAs for older adults’ health. They reported that CAs generally offer more benefits than challenges, such as ease of use and companionship, while identifying key challenges related to learning how to use CAs and ensuring privacy protection. Their review included studies involving both healthy older adults (\textit{N} = 18) and those with intellectual disabilities (\textit{N} = 5). However, it did not specify the types of CAs covered, and after examining its scope, it is evident that certain CA types, such as telephone-based interactive voice response systems, robots, and virtual agents without speech capabilities, were excluded.

Similarly, \citet{zhang_role_2024} conducted a scoping review of 29 papers to examine the use of chatbots in healthcare for older adults. Their work primarily described chatbot characteristics, evaluation approaches, and the purposes and challenges of applying chatbots in healthcare. They found that chatbots for older adults were primarily delivered through mobile applications with text-based interaction. While evaluations were generally positive, concerns about trust and operational difficulties persisted.

While these reviews provide valuable insights into the purposes, benefits, and challenges of CA applications \citep{even_benefits_2022, zhang_role_2024}, they fall short of offering a focused investigation of CAs from older adults' perspective. Specifically, they lack detailed exploration of the diverse forms and modalities of CAs, the lived experiences of older adults using these technologies, and their specific expectations for health-related applications. Addressing these gaps is essential for advancing the field. It enables HCI scholars to: (1) Develop a clear understanding of current CA designs, including their forms and modalities; (2) Align CA designs with older adults’ experiences and expectations, thereby identifying potential design practices for future innovations. This work aims to bridge these gaps by conducting a comprehensive systematic review of the research on CAs for older adults’ health. The review provides actionable insights into the design, evaluation, and implementation of CAs, supporting the development of more effective and user-centered technologies for older adults.

\section{Methods} \label{sec:methods}
\begin{figure}[h]
    \includegraphics[width=0.8\linewidth]{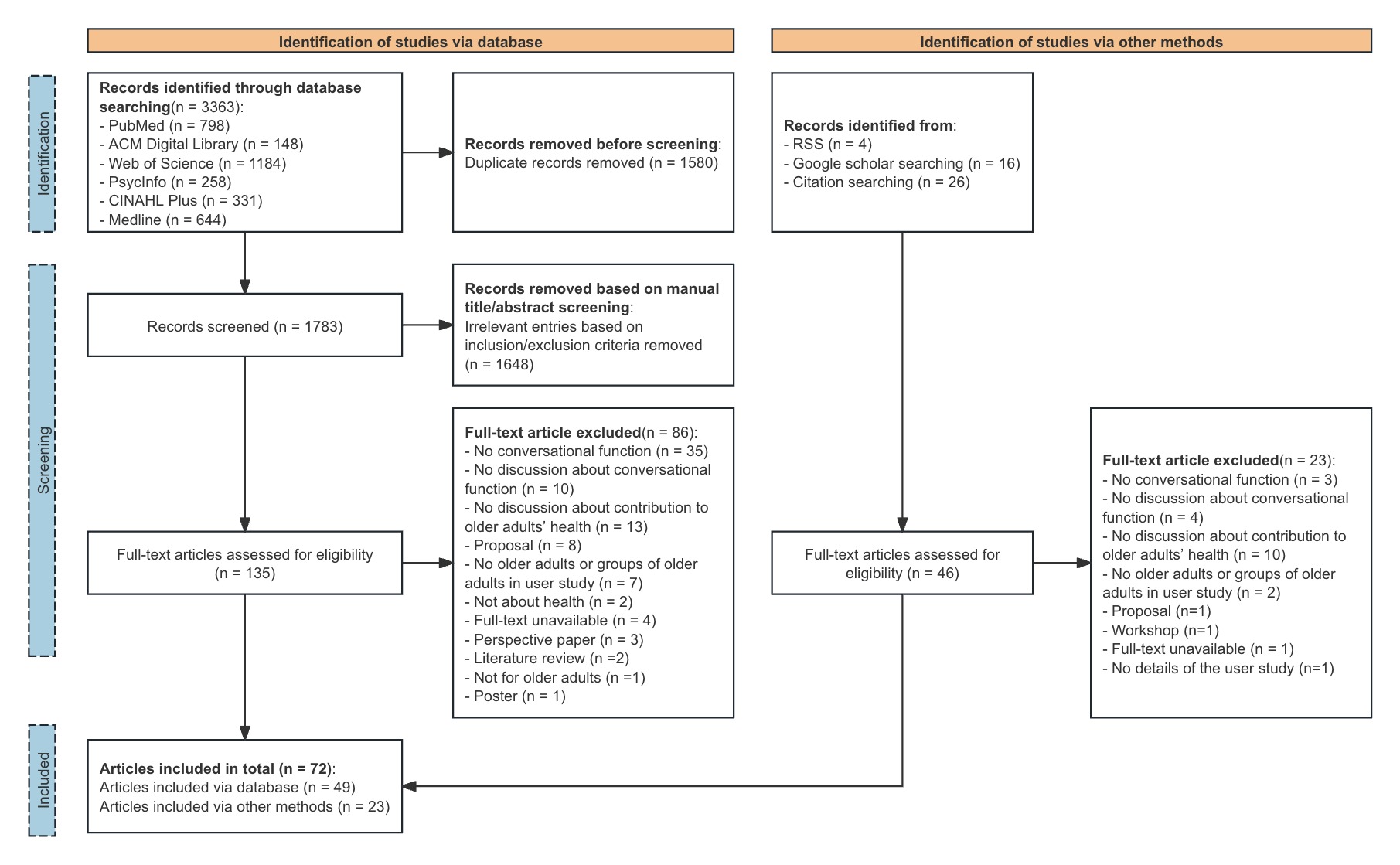}
    \caption{Flow diagram of our literature screening process}
    \label{fig:lite screening}
\end{figure}

We chose a systematic review method to gain a thorough understanding of CAs for older adults’ health and to guide future CA design. We adhered to the guidelines set forth by the Preferred Reporting Items for Systematic Review and Meta-Analysis (PRISMA) \cite{page_prisma_2021}. To outline a robust PRISMA process, in this section, we will present the literature review search (Section \ref{subsec:literature search}), review inclusion and exclusion (Section \ref{subsec:criteria}), study identification (Section \ref{subsec:study identification}), and the qualitative coding methodology (Section \ref{subsec:data extraction}). Figure \ref{fig:lite screening} demonstrates the whole literature screening process.

\subsection{Literature Search} \label{subsec:literature search}
We searched seven scholar databases: PubMed, Web of Science, PsycINFO, CINAHL Plus, Medline, ACM Digital Library, and Google Scholar. These databases were selected as they represent academic disciplines (e.g., medicine, psychology, human-computer interaction, and information science) that are most likely to explore CAs for older adults' health.
We then used an extensive list of search queries (see Appendix \ref{appendix:search queries}), including variations of health, older adults, and CA. These synonyms were generated by identifying specific terms or phrases used in article focused on older adults \cite{newman_social_2021} and CAs for health \cite{car_conversational_2020,laranjo_conversational_2018, el_kamali_virtual_2020}.
We searched "(synonyms of CA) AND (synonyms of older adults) AND (synonyms of health)" in the "title/abstract" field of PubMed and the "abstract" field of the rest databases except Google Scholar. For Google Scholar, the first author applied the search query "conversational agent AND older adults AND health" in "with all of the words" field of Google Scholar, screened the top 100 search result, and identified studies that were not included in previous databases. We didn't set the start year for search in order to depict CAs for older adults' health comprehensively. The first author conducted searches and screened out duplicates based on the title, abstract and doi number. The first search was conducted in May 2022, and the second search was conducted in June 2024 to update the literature. Additionally, we identified 30 relevant publications by tracking the references of the included articles and screening RSS of databases.

\subsection{Inclusion and exclusion criteria}  \label{subsec:criteria}
We developed the inclusion criteria based on the existing CA literature reviews \cite{zheng_ux_2022, car_conversational_2020, parmar_health-focused_2022, even_benefits_2022}. Only publications meeting the following criteria were included in this review: 
%A series of criteria were developed based on multiple rounds of discussions. We regarded "conversational function with natural language" as a key component of CAs based on the existing CA literature reviews\cite{zheng_ux_2022, car_conversational_2020, parmar_health-focused_2022,even_benefits_2022}, and aimed to include publications covering behavior study and prototype design. Researcher one (R1) and researcher two (R2) followed the criteria, screened 127 publications first, and reached an 88.98\% agreement rate. They then compared the screening results, addressed the differences, and revised the criteria with the third researcher (R3). Finally, they screened another 125 publications and reached a 96.83\% agreement rate. The final criteria are as follows.
\begin{itemize}
    \item Publications that included older adults as participants and reported the findings about them. As the age range of older adults in research communities differ \cite{lee_differences_2018}, we included studies that claimed older adults being included, i.e., the authors claimed in their title, abstract, or method that they focused on older adults or included older adults as a separate group of participants. For example, \citet{strasmann_age-related_2020} compared the attitudes towards the appearance of CA between older adults and young adults. Since their findings can provide insights for CA appearance design for older adults, we included their study in this review.
    \item The research context is older adults’ health. We included studies that explicitly claimed that they focused on older adults' health in title or abstract, or that investigated CAs in scenarios of health, such as medical care, mental/cognitive/physical illness, and everyday health and well-being. For example, \citet{tulsulkar_can_2021} didn't used the term "health" in their title, abstract and keywords, but they investigated the influence of CA on the cognitive and emotional well-being of cognitively impaired elderly. Thus, their paper was included in this review. 
    \item The research investigates the application of CAs, such as chatbots, robots with conversation functions, and voice assistants.
    \item Publications written in English that were published in peer-reviewed journals or conference proceedings.
\end{itemize}
%The previous review didn't include robots. However, human-robot interaction researchers are also interested in the \textit{embodied nature} of conversations, which is presented as a smart speaker or a humanoid robot\cite{mcmillan_human-robot_2023}. In this review, we also include the literature about robots and xxx.   

\subsection{Study identification}  \label{subsec:study identification}
% Fig. 1 shows the process of identifying eligible studies based on the inclusion/exclusion criteria. 
The first and second authors screened 127 publications first following the criteria. The agreement rate reached 88.98\%. We then compared the screening results and addressed the differences by consulting the third author. Finally, we screened an additional 125 publications and reached a 96.83\% agreement rate, indicating that the criteria were robust. The first and second authors subsequently screened the remaining articles, resulting in 97 articles for full-text review. The first author read the publications and further excluded 72 articles, leaving 25 papers included. In June 2024, the first and second authors repeated the search and screening process to update the literature, and 24 publications were finally added. An additional 23 articles were added through citation tracking and RSS. In total, we included 72 publications. As four publications \cite{bickmore_maintaining_2010,brewer_empirical_2022,ryu_simple_2020, santini_user_2021} include more than one user study, we overall collected 78 user studies.

\subsection{Data extraction}  \label{subsec:data extraction}
The data extraction process was conducted using Google Sheets. Key characteristics of the included articles were manually recorded, including publication venues, the geographical locations of publications based on authors' affiliations, and participant demographics. These demographics included the number of participants, age, gender, education, race, income, living conditions, and health conditions. The first author performed the initial data extraction, while the second author reviewed and validated a random sample comprising 33\% of the articles. No discrepancies were identified during this validation process.

For the categories "research method," "health domain," and study findings, the first author extracted the relevant key points as verbatim text into Google Sheets after a comprehensive review of the full-text articles. These extracted texts were subsequently analyzed using the qualitative content analysis approach \cite{miles_qualitative_2014}. An initial codebook was developed (see Table \ref{tab:codebook}), which categorized the data into: (1) research methods and health domains; (2) types and roles of conversational agents (CAs) (RQ1); (3) older adults’ experiences with CAs, including their effects, acceptance levels, and concerns regarding CAs (RQ2); and (4) older adults’ expectations of CAs (RQ3). The second author reviewed the extracted text and validated the coding results in accordance with the codebook. Any discrepancies were resolved collaboratively in group discussions.

\begin{table}[]
\caption{The codebook for data analysis.}
\setlength{\arrayrulewidth}{0mm}
\centering
\begin{tabular}{l|l|p{8cm}}
\toprule
\textbf{Category} &\textbf{Code} & \textbf{Definition}\\
\midrule
Research method& Qualitative & The research method that gather and analyse non-numerical data, e.g., interview and focus groups\\
 & Quantitative& The research method that gather and analyse numerical data, e.g., experiment and survey \\
 & Mixed method& The research design that combines quantitative and qualitative research methods to draw on the strengths of each\\
Health domains & General health& The studies that focus on older adults' general well-being, such as healthy aging\\
 & Mental health & The studies that focus on older adults' mental well-being and mental diseases, such as anxiety and depression.\\
 & Physical health & The studies that focus on older adults' physical well-being (e.g., exercise) and physical diseases (e.g., hypertension) \\
 & Cognitive health& The studies that focus on older adults' cognitive well-being (e.g., memory) and cognitive diseases (e.g., dementia) \\
Type of CAs& Physically ECA& The physically embodied CA, such as robots that can interact with human in natural languages\\
 & Graphically ECA & The CA that involves a virtual body, such as a human or animal avatar \\
 & Chatbot & The CA that only interact with users through text \\
 & Voice assistant & The CA that mainly interacts with users through voice, including commercial smart speakers (e.g., Google Home)\\
 Roles of CAs & Coach & To train older adults and help them adopt healthy habits and living skills\\
 & Companion & To be with older adults in their everyday life, help them alleviate mental health problems, and enhance their emotional status\\
 & General health assistant& To be with older adults in their everyday life and help with their everyday health and well-being \\
 & Therapist & To deliever therapy to older adults for mental illness treatment\\
 & Health information provider & To provide health information for older adults' learning and decision making\\
 & Monitor & To track and access older adults' symptoms and performance\\
 & Disease mangement assistant & To help older adults manage specific diseases \\
 Effect of CA & Level of effect & Positive, mixed, neutral, negative\\
 Acceptance of CA & Level of acceptance & Whether the findings indicate older adults' acceptance level of CAs, including three levels: low, neutral, and high \\
 Concerns of CA & Privacy & Older adults are concerned if they can be free from being observed by CAs and control the sharing of their personal data\\
 & Dependence& Older adults are worried about over-reliance on CAs \\
 & Cost& Older adults are afraid if they can afford a CA \\
 & Stigmatization& Older adults are concerned if they will be stigmatized (e.g., a weak person that can not live independently) because of using CAs \\
 & Usability & Older adults questioned CAs' ease of use and reliability\\
Expectations of CA & Personalization & Older adults should be able to adapt CAs to their personal preferences and needs, such as appearance and information service\\
 & Functionality & Older adults described their expected functions of CAs\\
 & Conversation & Older adults' expectation regarding the conversational style between CAs and them, including but not limited to voice, tone, length of message and so on\\
 & Appearance & Older adults' expectation regarding the appearance of CAs, such as gender, humanoid or not etc\\
 & Sense of control& Older adults expectation regarding their feelings of control over CAs\\ 
\bottomrule
\end{tabular}
\label{tab:codebook}
\end{table}

\section{Findings} \label{sec:findings}

To report the findings from our systematic review, we first present an overview of the descriptive statistics for the 72 publications included in this literature review (Section \ref{subsec:descriptive statistics}). We then summarize our coding themes to address each research question. Specifically for RQ1 (i.e., What are the CAs in older adults' health?), we report \textit{types of CA} and \textit{roles of CA} (Section \ref{subsec:CA for older adults}). For RQ2 (i.e., what are older adults experiences of using CAs for health?), we separately report the findings in three sections, first reporting \textit{the mixed effect of CA} on older adults' health, then describing older adults' \textit{low acceptance} and \textit{concerns} of using CAs (Section \ref{subsec:experiences of CAs}). Finally, we break down older adults' expectations of using CAs into\textit{ multiple functions}, \textit{natural conversational styles}, \textit{personalization}, and \textit{sense of control} (Section \ref{subsec:expectations of CAs}).

\subsection{Descriptive Statistics of Publications Reviewed} \label{subsec:descriptive statistics}

\subsubsection{Publication trend and venues}
Over the past 20 years, while there is a slight decrease in publications after 2022, the number of publications has increased overall. The indexed publications appeared in 58 different venues. We aggregated the publication venues that appeared only once (\textit{N} = 48) into "Others" in Table \ref{tab:pub_count} (See Appendix \ref{appendix:publication count} for the full list of publication venues). These venues covered all publications before 2015. Since 2016, The top 10 venues such as CHI, AMIA and JMIR had publications about "CAs for older adults' health", CHI is the most common one (7 papers). 
 
\subsubsection{Countries and regions}
As shown in Table \ref{tab:countries}, the publications are mainly from  North America (\textit{N} = 33), Asia (\textit{N} = 18), and Europe (\textit{N} = 13). Seven articles were from multi-country teams.

\begin{table} [ht]
    \caption{Countries of included studies}
    \setlength{\arrayrulewidth}{0mm}
    \centering
    \begin{tabular}{@{}l|p{7 cm}|}
    \toprule
        \textbf{Areas} & \textbf{Countries or regions} (\textit{N} of studies)\\
    \midrule
        North America & U.S. (32), Canada (1)\\
        Asia & Japan (10), Hong Kong (3), Taiwan (2), South Korea (2), Singapore (1)\\
        Europe & Italy (4), France (2), Germany (2), Netherlands (2), UK (2), Spain (1)\\
        Multiple countries & Germany, France, Italy, and Japan (2);  Austria, Italy, and the Netherlands (1); France, Norway, and Spain (1); Italy, Luxembourg, and Netherlands (1); Netherlands and Switzerland (1); UK and Japan (1)\\
        Oceania & Australia (1)\\
    \bottomrule
    \end{tabular}
    \label{tab:countries}
\end{table}

\subsubsection{Health domains}
We identified four main health domains from prior studies, including cognitive health (\textit{N} = 9), physical health (\textit{N} = 12), mental health (\textit{N} = 24), and general health (\textit{N} = 27). Table \ref{tab:health conditions} presents the four health domains and their corresponding health conditions or tasks. 
\begin{table} [ht]
    \caption{Three primarily health domains of prior studies focusing on CAs for older adults' health}
    \setlength{\arrayrulewidth}{0mm}
    \centering
    \begin{tabular}{@{}l|p{10cm}}
    \toprule
     \textbf{Health domains} & \textbf{Corresponding health conditions or tasks} (\textit{N} of studies) \\
    \midrule
        General health & General health assisting (13), Health information seeking (3), aging in place (2), Health topic learning (2), Diagnosis (1), Exercise (1), General health assessment (1), General health status monitoring (1), Health information seeking and medical contacting (1), Medication explanation (1), Mental health and fall prevention (1) \\
        Mental health & General mental well-being (5), Loneliness (5), Anxiety and depression (2), Reminiscence therapy (2), Self-disclosure (2), Depression (1), Loneliness and depression (1), Memory training (1), Self-reflection (1), Social activities (1), Social isolation (1), Social support (1), Stress and anxiety (1) \\
        Physical health & Exercise (6), Diet (2), Diabetes (1), Diabetes and Hypertension (1), Hypertension (1), Type 2 Diabetes (1) \\
        Cognitive health & Dementia (6), Cognitive training (1), Mild cognitive impairment (1), Neurocognitive disorders (NCD) screening (1) \\
    \bottomrule
    \end{tabular}
    \label{tab:health conditions}
\end{table}

\begin{figure} [ht]
    \centering
    \includegraphics[width=0.8\linewidth]{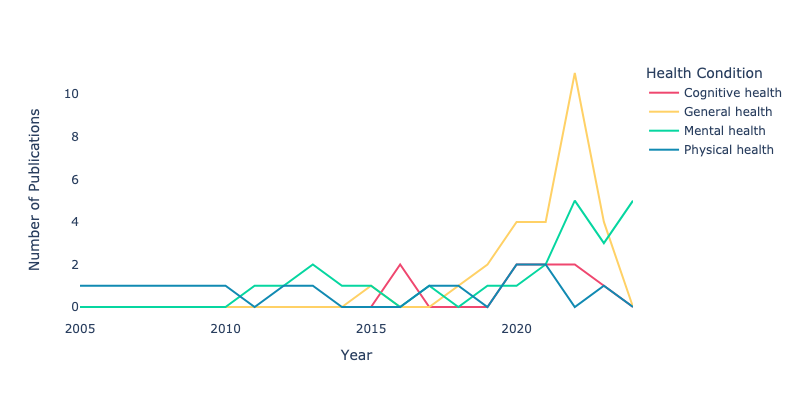}
    \caption{Different health research domains over years}
    \label{fig:health domains}
\end{figure}

We then depicted the publication trend in Figure \ref{fig:health domains}. According to the figure, we found that while research have initially focused on physical health, researchers gradually broadened their scope to include mental health and general health.

\subsubsection{Research method}
Among these 72 included papers, research methods they adopted show an equal distribution with 28 of them using quantitative, 20 of them using qualitative, and 24 of them applying mixed method approaches. Among the articles adopted the quantitative approach, 26 used an experimental design: five of them were conducted in the field (e.g., the eldercare home \cite{tulsulkar_can_2021}) and two of them applied Wizard-of-Oz; the experiment time varies from 27.3 min on average \cite{azevedo_using_2018} to 388 days \cite{lima_discovering_2023}. One publication used questionnaires to explore older adults evaluation of a voice-based interface for diabetes \cite{cheng_development_2018}. Another publication investigated older adults' engagement with a digital coaching program by log analysis \cite{graham_older_2021}. Regarding measurement, the studies either measured participants' task performance (e.g., conversation times \citep{graham_older_2021}) or applied multiple existing or self-defined scales. The commonly used scales included the UCLA loneliness scale (\textit{N} = 6), Geriatric Depression Scale (\textit{N} = 4), World Health Organization-Quality of life questionnaire (\textit{N} = 3), system usability scale (N = 4), and Patient Health Questionnaire-8 (PHQ-8) \footnote{The eight-item Patient Health Questionnaire depression scale (PHQ-8) is established as a valid diagnostic and severity measure for depressive disorders.} (\textit{N} = 2).
 
Regarding the qualitative approach, the included articles applied methods including semi-structure interviews (\textit{N} = 14), focus groups (\textit{N} = 4), diary (\textit{N} = 2), observation (\textit{N} = 2), and workshops (\textit{N} = 2). Several studies applied longitudinal design, ranging from one week \cite{vardoulakis_designing_2012} to three months \cite{chi_pilot_2017} to examine the long-term effect of CAs on older adults. For papers applying mixed method approach, a common pattern is to combine experimental studies with semi-structured interviews (\textit{N} = 14). 

\subsubsection{Participant characteristics}
Table \ref{tab:p_characteristics} provided an overview of participant characteristics in 78 studies reported in the 72 included articles (See Appendix \ref{appendix:p_characteristics}). The sample size of older adult participants ranged from 3 to 301 (\textit{M} = 38.53, \textit{SD} = 57.31). Most of these studies (\textit{N} = 65) involved fewer than 50 older adult participants. We described participant characteristics as follows:

\textbf{Demographics. }
Across the 78 studies from 72 included articles, there were significant variations in the age ranges of the older adult participants. This might reflect differing definitions of aging populations across countries.

Specifically, 45 studies reported the age range of participants, the minimum age was between 50 and 77 (\textit{M} = 62.93, \textit{SD} = 6.40) and the maximum age was between 67 and 98 (\textit{M} = 84.57, \textit{SD} = 8.15). 42 studies reported the mean age of participants, which is between 55.58 and 81.90 (\textit{M} = 71.17, \textit{SD} = 6.70); Six studies did not report age ranges or mean values but defined older adults varingly as people above 50 \cite{miura_assisting_2022}, 55 \cite{ring_addressing_2013, wang_promoting_2024, jin_exploring_2024}, 60 \cite{striegl_investigating_2022}, and 65 years old \cite{justo_analysis_2020}.
Six studies claimed that they recruited older adults as participants without specifying the age threashold \cite{ryu_simple_2020, tulsulkar_can_2021, betriana_interactions_2021, cheng_development_2018, baskar_human-agent_2015}.

The included studies also mainly recruited participants who were
(1) \textit{Female}: Of the 78 studies, 66 reported participants' gender. The average percentage of female participants is 65.99\% (\textit{SD} = 18.38\%). 89.39\% studies (\textit{N} = 59) had a sample with more than 50\% being women participants. 
(2) \textit{Highly-educated}: About one-third of the studies reported participants' education (\textit{N} = 27), and their average percentage of participants with high school and lower education was 43.73\% (\textit{SD} = 30.06\%); only 37.04\% studies (\textit{N} = 10) recruited more than 50\% of participants with high school and lower education; 
(3) \textit{Multiracial}: About one third of studies reported participants' races (\textit{N} = 27), including multi-races (\textit{N} = 18), Asian (\textit{N} = 6), Black or African American (\textit{N} = 1), and White or Caucasian (\textit{N} = 1). Studies with multiracial participants recruited participants in White or Caucasian (\textit{N} = 19), Black or African American (\textit{N} = 14), Asian (\textit{N} = 17), Hispanic or Latino (\textit{N} = 4), Native American (\textit{N} = 2), American Indian or Alaska Native (\textit{N} = 1), and Native Hawaiian or Pacific Islander (\textit{N} = 1). 

Only seven studies reported the income conditions of the participants, and three of them focused on low-income older adults with different criteria, including annual income below \$35,000 \cite{mccloud_using_2022} or \$20,000 \cite{nallam_question_2020}, and whether the participants lived in low-income housing \cite{chung_smart_2023}.
%One study didn't reported participants' education background but indicated that 19 out of 21 participants had low reading literacy \cite{bickmore_its_2005}. 

\textbf{Health conditions. }
Nearly half of the studies (\textit{N} = 38) reported participants' health status. We grouped them into the following categories: 
(1) Older adults with health conditions (\textit{N} = 14). This group of research mainly recruited older adults with cognitive or mental conditions such as dementia (e.g.,  \cite{stara_toward_2021, betriana_interactions_2021, stara_usability_2021}), depression (e.g., \cite{dino_delivering_2019, khosla_embodying_2013}) and other mild cognitive impairment (e.g., \cite{lima_discovering_2023, wargnier_field_2016}). Only two studies recruited older adults with physical health problems such as malnutrition \cite{kramer_use_2022} and diabetes \cite{chou_user-friendly_2024}. 
(2) Older adults who were healthy in specific aspects (\textit{N} = 9). This group of study mainly held the criteria that participants should have no mental illness (e.g., \cite{ter_stal_embodied_2020, bickmore_randomized_2013}) or have good hearing and vision \cite{cuciniello_cross-cultural_2022} to guarantee that participants were able to interact with CAs.
(3) Older adults with different levels of health status (\textit{N} = 8), e.g., \cite{kanoh_examination_2011} recruited both healthy older adults and older adults with cognitive impairments; 
(4) Healthy older adults (\textit{N} = 7).

\textbf{Living condition. }
Thirty-four studies reported participants' living conditions. Studies with participants in multiple living status (\textit{N}= 9) recruited older adults who lived alone (\textit{N}= 8), lived with others (e.g., caregivers and families) (\textit{N}= 6), or lived in care facilities (\textit{N}= 2), communities (\textit{N}= 1), and independent household (\textit{N}= 1). Two studies also applied the Activities of Daily Living (ADL) scale to measure participants' independent living ability. Six studies recruited older adults who lived with others, including partners, families, and caregivers.

%\subsubsection{Experience of using CAs}
%20 studies reported participant's experience with CAs or other technologies and 15 of then recruited older adults with different levels of experience using CAs. 
\subsection{Type and roles of CAs for older adults' health} \label{subsec:CA for older adults}
We analyzed the exploration of CAs for older adults' health in previous research from the following aspects: (1) Types of CAs. We categorized CAs into the following types based on their modality and embodiment: chatbot, voice assistant, graphically ECA, and physically ECA. And we noticed an increasing interest on chatbots and voice assistants. (2) Roles of CAs. We identified the following roles that CAs play for older adults health in previous research: companion, coach, general health assistant, therapist, health information provider, monitor, and disease management assistant.

\subsubsection{Types of CAs}\label{subsubsec:types of CAs}
\begin{figure}
    \centering
    \includegraphics[width=0.75\linewidth]{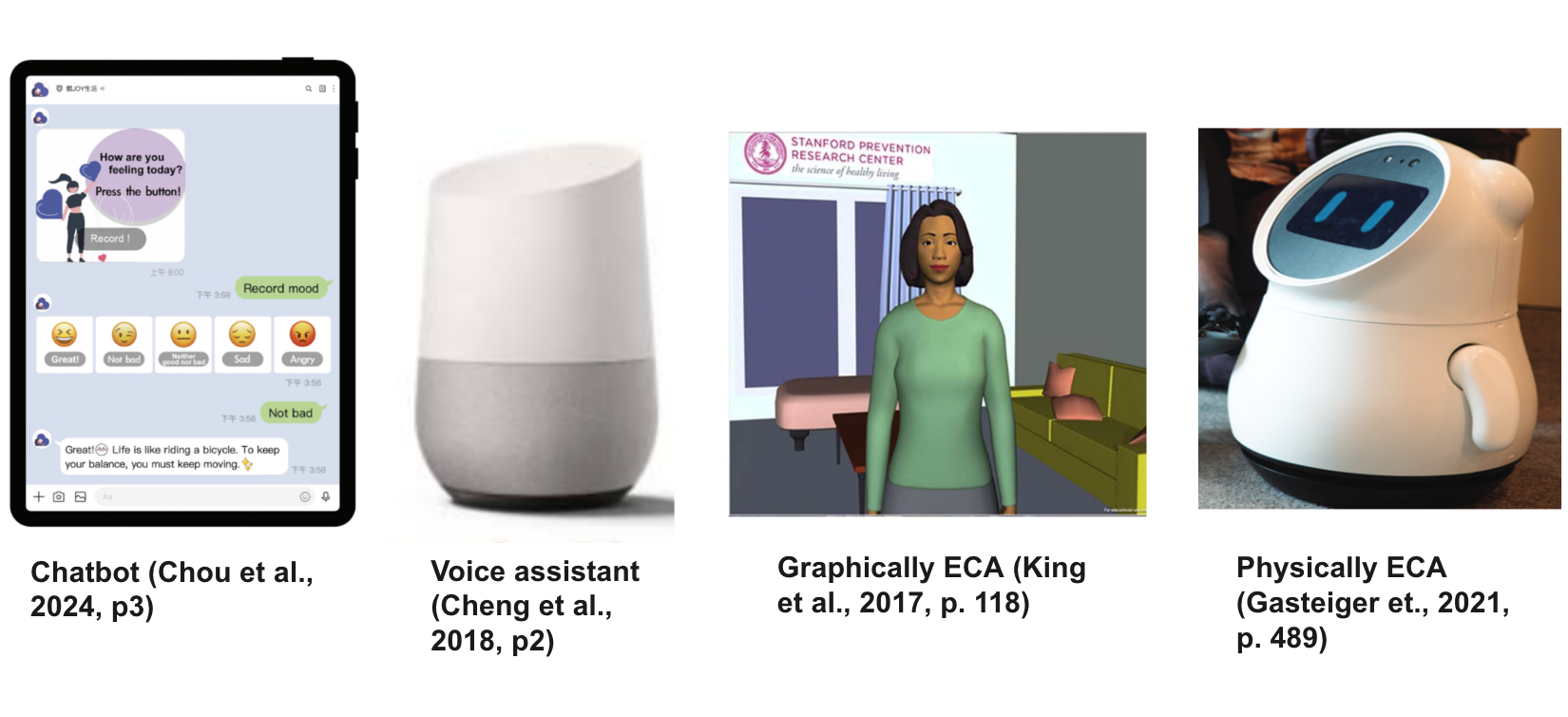}
    \caption{The examples of four types of CAs: Chatbot \cite{chou_user-friendly_2024}, Voice assistant \cite{cheng_development_2018}, Graphically ECA \cite{king_testing_2017}, and Physically ECA \cite{gasteiger_older_2022}}
    \label{fig:CA example}
\end{figure}

% Following the categories from the previous literature review \cite{allouch_conversational_2021, montenegro_survey_2019}, 
We identified four broad types of CAs from the included articles (see fig. \ref{fig:CA example} for examples). This category is derived from a review of existing literature on CAs in general \citep{allouch_conversational_2021, montenegro_survey_2019}. First, a \textit{chatbot} represent a particular type of CA that only interact with users through text, such as ELIZA chatbot and chatbots in service platforms (e.g., The Lark health platform \cite{graham_older_2021}). Second, a \textit{voice assistant} generally refers to CA that mainly interacts with users through voice, including commercial smart speakers (e.g., Google Home) and self-developed prototypes. Some voice assistants would integrate GUI interaction. Third, a \textit{graphically ECA} typically involves a virtual body, such as a human or animal avatar. Finally, a \textit{physically ECA} is a type of CA that is physically embodied, such as social robots, and can interact with humans in-person.

As shown in Table \ref{tab:CA_type}, we found that nearly half of the publications (\textit{N} = 39) investigated graphically ECAs and physically ECAs. For voice assistants, research mainly focused on commercial smart speakers, including Amazon Alexa \cite{brewer_if_2022}, Amazon Echo \cite{yan_impact_2024}, Amazon Echo Show \cite{soubutts_amazon_2022, jones_reducing_2024, lima_discovering_2023}, Amazon Echo Dot \cite{jones_reducing_2024}, Google Home \cite{harrington_its_2022}, and Google Home Mini \cite{brewer_empirical_2022, desai_ok_2023, desai_painless_2023,chin_user_2020}. For chatbots, several publications developed prototypes based on open source or commercial platform, including Google Cloud Dialogflow \cite{valtolina_charlie_2021, griffin_chatbot_2023}, LINE \cite{miura_assisting_2022, chou_user-friendly_2024}, Kakao Developer \cite{ryu_simple_2020}, and Amazon Alexa \cite{striegl_investigating_2022}; one publication specifically investigated the chatbot of commercial healthcare platform Lark Health \cite{graham_older_2021}.

\begin{table} [ht]
    \caption{Taxonomy of different CAs based on forms and modality. "User need investigation" means “understanding existing user needs for design”.}
    \setlength{\arrayrulewidth}{0mm}
    \centering
    \begin{tabular}{@{}lll}
    \toprule
        \textbf{Category I} & \textbf{Category II} & \textit{N} of studies \\
    \midrule
        Graphically ECA & Graphically ECA & 23 \\
         & User need investigation & 1 \\
        Physically ECA & Robot & 15 \\
        Voice assistant & Commercial smart speaker & 13 \\
         & Self-developed prototype & 5 \\
         & User need investigation & 3 \\
        Chatbot & Self-developed prototype & 9 \\
         & Commercial Chatbot & 1 \\
         & User need investigation & 1 \\
        Voice assistant + Chatbot & Self-developed prototype & 1 \\
    \bottomrule
    \end{tabular}
    \label{tab:CA_type}
\end{table}

\begin{figure} [ht]
    \centering
    \includegraphics[width=0.8\linewidth]{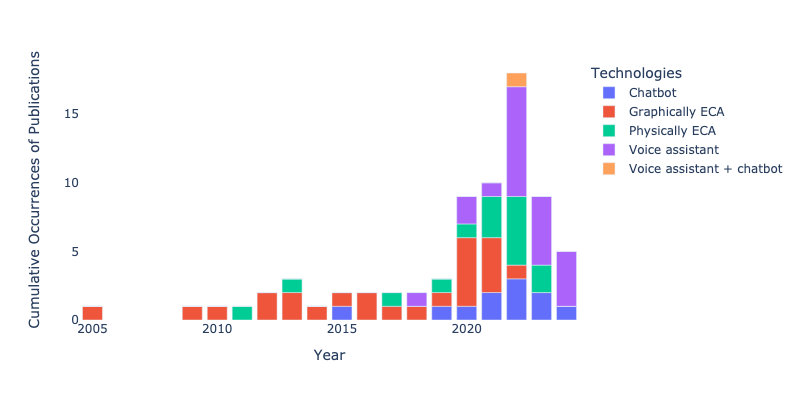}
    \caption{Different types of CAs studied over years}
    \label{fig:year*CA type}
\end{figure}

As shown in Figure \ref{fig:year*CA type}, earlier research mainly focuses on Graphically ECAs. Since 2019, the number of publications has significantly increased with a shift of focus from Graphic ECAs to chatbots and voice assistants. We also noticed one publication that designed a ChatGPT-powered conversational companion for older adults and obtained positive evaluations from experts \cite{alessa_towards_2023}. As this publication is not included in this review due to its lack of older adult participants, it shows the trend that researchers are trying to empower CAs with generative AI for older adults' health. 

\subsubsection{Roles of CAs} \label{subsubsec:roles of CAs}
%As \citet{Khadpe2020-bi} found that different portrayals of AI significantly influence how competent people perceive these AI robots to be, which in turn affects their adoption.
%Note that the types of CAs refer to its forms and modality (e.g., text, voice, or multi-modal), while its The role of CAs refers to the purpose, functionalities, or metaphorical definitions that researchers or older adults perceive CAs.  
We identified seven main roles that CAs play based on the included articles.

(1) \textit{Companion} (\textit{N} = 17). This research applied CAs as companions to help alleviate older adults' loneliness \cite{valtolina_charlie_2021, ring_social_2015, ring_addressing_2013, chi_pilot_2017, chung_smart_2023, jones_reducing_2024, yan_impact_2024}, improve their social activity \cite{kanoh_examination_2011}, or enhance their emotional status \cite{khosla_embodying_2013, tsujikawa_development_2023}. These CAs were mainly physically ECAs (\textit{N} = 6), graphically ECAs (\textit{N} = 6), and commercial smart speakers (\textit{N} = 3). As companions, the CAs were designed with functions such as entertainment (e.g., playing quizzes \cite{kanoh_examination_2011, valtolina_charlie_2021}) and proactive conversations (e.g., initiating conversations by detecting older adults' walk \cite{ring_social_2015}).

(2) \textit{Coach} (\textit{N} = 16). This research applied CA as coaches for exercise \cite{obo_human-robot_2017, betriana_interactions_2021, bickmore_its_2005, bickmore_virtual_2009, bickmore_randomized_2013, bickmore_maintaining_2010, chin_user_2020}, healthy aging \cite{santini_digital_2020, santini_user_2021, mctear_interaction_2023, moller_user_2022, cuciniello_cross-cultural_2022}, diet \cite{justo_analysis_2020, kramer_use_2022}, managing diabetes and hypertension \cite{graham_older_2021}, and conversational skill training \cite{ali_aging_2021}. The CAs are mainly graphically ECA (\textit{N} = 10). Two studies combined psychological and behavioral theories in guiding the design of CAs. Based on the cognitive behavioral theory, \citet{graham_older_2021} created regular "nudges" and "call-to-action" to encourage users to engage in the interaction with CAs, such as having a coach conversation or positively reinforce users (e.g., great job on your walk today). \citet{justo_analysis_2020} designed the dialog content following the GROW model, including four phases: Goals or objectives, Reality, Options, and Will or action plan.

(3) \textit{General health assistant} (\textit{N} = 12). This research targeted general health, such as general well-being \cite{papadopoulos_caresses_2022, mccloud_using_2022, soubutts_amazon_2022}, healthy aging \cite{stara_toward_2021}, aging in place \cite{leung_exploring_2023}, and quality of life \cite{chen_understanding_2021}. These CAs included voice assistants (\textit{N} = 5), graphically ECAs (\textit{N} = 3), physically ECAs (\textit{N} = 2), and chatbots (\textit{N} = 1). They could include multiple functions, such as talking as a companion \cite{papadopoulos_caresses_2022, stara_usability_2021, bott_protocol-driven_2019, leung_exploring_2023}, providing medication reminders \cite{stara_toward_2021, stara_usability_2021}, contacting healthcare providers \cite{bott_protocol-driven_2019}, tracking symptoms \cite{chou_user-friendly_2024}, health education \cite{chou_user-friendly_2024}, and entertainment \cite{papadopoulos_caresses_2022, stara_toward_2021, stara_usability_2021, bott_protocol-driven_2019, leung_exploring_2023}.

(4) \textit{Therapist} (\textit{N} = 9). This research used CAs to deliver therapies to treat mental illness such as anxiety \cite{danieli_assessing_2022} and depression \cite{park_impact_2022, dino_delivering_2019} or improve older adults' emotional well-being \cite{nakamura_effect_2022, wang_promoting_2024, jin_exploring_2024, randall_realizing_2023} and cognitive ability \citep{sugimoto_tract-based_2022, striegl_investigating_2022}. These CAs were mainly physically ECAs (\textit{N} = 5) and voice assistants (\textit{N} = 1). Some studies based CA designs on existing therapy methods, such as the cognitive behavior therapy \cite{dino_delivering_2019, striegl_designing_2021} and the reminiscence therapy \cite{jin_exploring_2024, randall_realizing_2023}. 

(5) \textit{Health information provider} (\textit{N} = 9). This research applied voice assistants (\textit{N} = 5) or chatbots (\textit{N} = 1) to help older adults obtain health information \citep{wu_acceptance_2019, brewer_empirical_2022, brewer_if_2022, desai_ok_2023, harrington_its_2022, nallam_question_2020}.  

(6) \textit{Monitor} (\textit{N} = 6). This research used CAs to assess older adults health status, such as neurocognitive disorders screening \cite{ding_talktive_2022} and mental well-being of older adults with dementia \citep{lima_discovering_2023}. These CAs included chatbots (\textit{N} = 3), graphically ECAs (\textit{N} = 2), and voice assistants (\textit{N} = 1).

%Engagement with Theories or Theoretical Framework:
%21 out of 74 included publications engaged with theories or theory-based concepts to frame their study in the designing of a prototype or working system. We elaborate on such engagement based on the types of roles that CA plays in older adults' health:

(7) \textit{Disease management assistant} (\textit{N} = 3). This research applied CAs to help older adults manage specific diseases, including a Google home application for older adults with type 2 diabetes to collect their data and generate suggestions \cite{cheng_development_2018}, a chatbot supporting older adults' management of hypertension \cite{griffin_chatbot_2023}, and a chatbot helping older adults deal with anxiety and depression \cite{ryu_simple_2020}.

Table \ref{tab:CA type and role} shows the interaction between CA type and roles. CAs as companions, coaches, and therapists were mainly in the forms of Graphically or Physically ECAs (70.59\% for companions, 62.5\% for coaches, 55.56\% for therapists); CAs as health information providers mainly in the form of voice assistants.
% \usepackage{graphicx}
\begin{table}[]
    \caption{The cross table between types of CAs and roles of CAs }\label{tab:CA type and role}
    \centering
    \setlength{\arrayrulewidth}{0mm}
    \begin{tabular}{@{}lp{1.5cm}p{1.5cm}p{1.3cm}lp{2.4cm}c}
    \toprule
% & \multicolumn{5}{c}{\textbf{Types of CAs}} &  \\
\textbf{} & \textbf{Graphically ECA} & \textbf{Voice assistant} & \textbf{Physically ECA} & \textbf{Chatbot} & \textbf{Voice assistant + chatbot} & \textbf{Total} \\
    \midrule
\textbf{Companion} & 6 & 4 & 6 & 1 & 0 & 17 \\
\textbf{Coach} & 10 & 2 & 2 & 2 & 0 & 16 \\
\textbf{General health assistant} & 4 & 5 & 2 & 1 & 0 & 12 \\
\textbf{Health information provider} & 2 & 6 & 0 & 1 & 0 & 9 \\
\textbf{Therapist} & 0 & 2 & 5 & 1 & 1 & 9 \\
\textbf{Monitor} & 2 & 1 & 0 & 3 & 0 & 6 \\
\textbf{Disease management assistant} & 0 & 1 & 0 & 2 & 0 & 3 \\
\textbf{Total} & 24 & 21 & 15 & 11 & 1 & 72\\
    \bottomrule
\end{tabular}%
\end{table}

\subsection{Older Adults' Experiences of CAs}  \label{subsec:experiences of CAs}
Among the 78 included studies, 36 reported the effect of CAs on older adults' health and 26 investigated older adults' attitudes towards CAs. We found a mixed effect of CAs on older adults' health. Older adults generally showed a low acceptance of CAs and had some major concerns of CAs. 

\subsubsection{Mixed effect of CAs on older adults' health} \label{subsec:effects of CAs}
Thirty-six studies have reported the effects of CAs on older adults after they used CAs for certain health activities. The results included 26 positive, 8 neutral, and 2 negative effects. For positive effects, the included papers observed that CAs can encourage older adults to walk more steps (\textit{N} = 4), help them maintain social connections (\textit{N} = 3), keep healthy diet (\textit{N} = 2), and better support their cognitive ability (\textit{N} = 2) or emotional well-being (\textit{N} = 8). The positive effects of CAs on older adults' mental health included significantly decreased anxiety \cite{ryu_simple_2020}, depression\cite{danieli_assessing_2022} and loneliness \cite{park_impact_2022, bott_protocol-driven_2019}. However, only five studies \cite{mccloud_using_2022, park_impact_2022, bickmore_its_2005, bickmore_randomized_2013, miura_assisting_2022} grounded their results on a longitudinal (more than two months) observation. Among these papers, \citet{bickmore_randomized_2013} reported that the positive effect of CAs on older adults' walking steps did not last long: Older adults who used conversational agents (CAs) as coaches experienced a significant increase in walking steps at first, but this effect was no longer significant by 12 months. This might be due to the novelty effect, where older adults might be curious about CAs and use them actively at first, and their frequency of use would decrease when curiosity wanes \cite{lima_discovering_2023}. 

Meanwhile, there were also studies (\textit{N} = 8) that did not observe significant effects of CAs on older adults' activity \cite{tulsulkar_can_2021}, physical health \cite{papadopoulos_caresses_2022}, diet \cite{kramer_use_2022}, exercise \cite{bickmore_maintaining_2010}, or emotional well-being \cite{papadopoulos_caresses_2022, kramer_use_2022, danieli_assessing_2022, bickmore_its_2005}. Possible contributing factors included the types and roles of CAs, intervention time, and measurements, as well as environmental changes. For example, \citet{danieli_assessing_2022} argued that the effect of CAs on older adults' anxiety levels may bediminished by environmental changes such as the increase of COVID-19 positive cases and a general concern arising from new virus variants. 

Two papers \cite{stara_toward_2021, soubutts_amazon_2022} reported negative effects of CAs. \citet{stara_toward_2021} found that some older adults with dementia held a contradictory attitude towards CAs: while they thought that CAs could help improve their confidence of living, they felt that the existence of CAs constantly reminded them of the fact "getting olders" \citet{soubutts_amazon_2022} observed the increased social tension among family members due to the conflict of using Amazon Echo. An example was that older adults' interaction with Echo disturbed the activities of other family members.

\subsubsection{Low acceptance of CAs} \label{subsubsec:acceptance of CAs}
Among 15 studies that reported older adults' acceptance level of CAs, 11 found that older adult have a low acceptance of CAs. Some studies found that older adults preferred human assistance or using their familiar technologies (e.g., laptops) rather than CAs in the activity program \cite{kanoh_examination_2011}, health information seeking \cite{brewer_empirical_2022}, conversational skill coach \cite{ali_aging_2021}, and social companion \cite{vardoulakis_designing_2012}. However, older adults could become more receptive to CAs when they were informed of CAs' functions and potential benefits \cite{santini_digital_2020, santini_user_2021}, especially when the functions fit in their specific needs, such as learning health knowledge \cite{wu_acceptance_2019}, preventing dementia \cite{ryu_simple_2020}, and obtaining emotional support \cite{striegl_investigating_2022}. 

Several studies also explored factors that impact older adults' acceptance of CAs. They did not find the influence of gender, age, or experience with technology \cite{tsiourti_virtual_2014}, but identified other factors that discouraged older adults' CA acceptance, such as the stigma of using CAs \cite{santini_digital_2020, ryu_simple_2020, ring_social_2015} and older adults' interaction fluency with CAs \cite{desai_ok_2023, harrington_its_2022}. 

For stigma, older adults seem to regard using CA as a sign of "someone who needs help." For example, \citet{santini_digital_2020} and \citet{ring_social_2015} found that older adults would differentiate themselves from other older adults who suffered from mental health problems, physical disabilities, or cognitive impairments, commenting that CAs would be more useful for the latter group. To avoid being labeled as "someone who needs help", older adults would not use CAs \cite{ryu_simple_2020}. 

Regarding the interaction fluency, \citet{desai_ok_2023} defined it as the inverse of the number of breakdowns or failures that occur during conversations between older adults and the CA. They found that older adults' acceptance of CAs was more influenced by interaction fluency than younger adults. \citet{harrington_its_2022} identified factors that impacted black older adults' interaction fluency with voice assistance: their unfamiliarity with voice interaction and CAs' inability to understand dialects and accents, and thus negatively affected their acceptance of CAs. 

%- Older adults' evaluation of CA
\subsubsection{Concerns of CAs} \label{subsubsec:concerns of CA}
Nine studies identified older adults' concerns of using CAs \cite{tsiourti_virtual_2014, santini_user_2021, chen_understanding_2021, vardoulakis_designing_2012, chi_pilot_2017, chung_smart_2023, nallam_question_2020, moller_user_2022, jin_exploring_2024}. All of them placed \textbf{privacy} as a key issue in accepting and trusting CAs, which is mainly about personal data protection. For example, \citet{chen_understanding_2021} interviewed 16 older adults about their opinions of voice assistants in healthcare. They found that older adults were convinced that their speech data would be illegally leaked and were concerned about the transparency of data storage, use, and sharing. When the privacy concern was eliminated through researchers' introduction of CAs' functions, \citet{jin_exploring_2024} found that older adults were more willing to share their data with CAs. 

The second main concern was about \textbf{dependence}, either physically or mentally \cite{santini_user_2021, chi_pilot_2017, chung_smart_2023, moller_user_2022, jin_exploring_2024}. For example, \citet{santini_user_2021} investigated user requirements of an ECA for healthy aging through focus groups. They found that older adults were afraid of being dependent on CAs, letting CAs substitute their cognitive and decision-making capabilities. \citet{chung_smart_2023} found similar concerns from live-alone older adults, who were afraid of getting used to fast information from CAs would reduce their motivations to think critically and learn new things. Apart from cognitive dependence, some older adults were also concerned about dependence on social and physical activities. \citet{chi_pilot_2017} conducted a pilot test of a digital pet avatar with older adults. Although older adults enjoyed the avatar's companion, they expressed concerns over its substitution of their interaction with real-world humans. In the study conducted by \citet{santini_user_2021}, older adults had concerns regarding their over-reliance on CAs to finish all tasks (e.g., bring me the slippers), resulting in their inactivity and the loss of controlling of their life.

Others concerns reported in the included studies included \textbf{cost} \cite{chi_pilot_2017, moller_user_2022}, \textbf{stigmatization} \cite{santini_user_2021}, \textbf{usability} \cite{nallam_question_2020}, and \textbf{information quality} \cite{nallam_question_2020}. \citet{chi_pilot_2017} and \citet{moller_user_2022} reported that older adults worry about the cost of electricity, Internet, and CA itself and that these factors would discourage them from adopting CAs. For stigmatization, \citet{santini_user_2021} pointed out that older adults would feel "being stigmatized as old" because they interpreted the need of using CAs as a sign of "being old and not self-sufficient".  \citet{nallam_question_2020} reported older adults' questions about whether the medical information from CAs was up-to-date and trustworthy and CAs' availability when the Internet was cut off.

\subsection{Older Adults' Expectations of CAs} \label{subsec:expectations of CAs}
Twenty-three studies focused on older adults' needs for and expectations of CAs in their healthcare. We identified four main themes, including functionalities, natural conversations, personalization, and a sense of control.

\subsubsection{Functionalities} \label{subsubsec:funtionalities}
In 13 studies, older adults expected that CAs could play multiple roles in their daily healthcare, such as activity management \cite{tsiourti_virtual_2014,santini_user_2021, lifset_ascertaining_2023}, medication management \cite{tsiourti_virtual_2014, lifset_ascertaining_2023}, emergency assistance \cite{chung_smart_2023, nallam_question_2020}, mental support \cite{santini_digital_2020, lifset_ascertaining_2023, tsiourti_virtual_2014, santini_user_2021, chung_smart_2023, moller_user_2022}, and information collection \cite{harrington_its_2022, nallam_question_2020,soubutts_amazon_2022, santini_digital_2020, santini_user_2021, brewer_empirical_2022}. For activity and medication management, \citet{tsiourti_virtual_2014} interviewed 20 older adults from the Netherlands and Switzerland and noticed older adults' proneness to forgetfulness. To accomplish their daily routine activities, older adults hoped that CAs could maintain a personal up-to-date daily plan, monitor activity execution, and decide about issuing appropriate reminders for basic actions (e.g., taking medication). Emergency assistance was also essential for older adults, especially those who live alone. \citet{chung_smart_2023} conducted focus groups with 25 live-alone older adults. They found that the most important reason for older adults to adopt smart speakers was to get help in emergency situations, such as falling. \citet{nallam_question_2020} also identified the need for emergency assistance from older adults who live alone.

Several papers focused on older adults' needs for mental support in different contexts. \citet{lifset_ascertaining_2023} interviewed 16 older adults during the COVID-19 pandemic and found that 10 of them mentioned isolation, depression, and anxiety as substantial health concerns. CAs may provide companionship through personalized interactions to help address these concerns. \citet{santini_digital_2020} focused on older adults' retirement; they interviewed 10 older workers and retirees from Italy and noticed the emotional discomfort during the transition from work to retirement due to the potential loss of social relationships and cognitive stimuli. Participants suggested CA functions to help retirees be prepared for retirement, such as enhancing retirees' self-awareness, helping them to identify their needs and desires, and connecting them by building communities. \citet{tsiourti_virtual_2014} found that older adults regarded emotional companion from CAs as a motivator of undertaking activities that are good for their health, such as physical exercises. \citet{chung_smart_2023} found that live-alone older adults expected CAs to help them connect with others in social activities, such as going for a walk or grocery shopping, tailored their preference.
 
Older adults expressed needs for CAs that could help them access and manage health information across different topics, such as illness and medication \cite{harrington_its_2022, nallam_question_2020}, lab work \cite{harrington_its_2022}, COVID \cite{soubutts_amazon_2022}, diet and nutrition \cite{harrington_its_2022}, or address aging-related topics, such as retirement \cite{santini_digital_2020, santini_user_2021}; \citet{desai_painless_2023} conducted a semi-structured interview and co-design activities with 10 older adults to develop a voice-based user interface (VUI) that can engage older adults in informal health information learning. They found that older adults expected to know the source of all health information, select their preferred sources, and obtain resources for further learning. \citet{brewer_empirical_2022} conducted a survey of 201 older adults to understand their health information-seeking with Google Home Mini, finding that older adults expected detailed responses, personalized health information, and recommendations from Google.

\subsubsection{Natural conversations} \label{subsubsec:expect natural conversations}
Nine studies examined older adults' anticipated conversation style when interacting with CAs. Overall, they generally preferred more natural and fluent conversations with CAs, in a way similar to human-human communications \cite{brewer_empirical_2022, mctear_interaction_2023, moller_user_2022, jin_exploring_2024, ter_stal_embodied_2020, tsiourti_virtual_2014, santini_user_2021}. For voice tone, they expected that CAs could adapt their voices to the conversation context \cite{moller_user_2022}, e.g., voice should match the emotions in the conversation \cite{jin_exploring_2024}. In terms of conversation content, they hoped that CAs could engage in longer dialogues so that users could introduce new topics freely rather than answering questions by following a sequence of orders \cite{mctear_interaction_2023}.  Additionally, they expected that CAs would use different conversational styles depending on the context. For example, they expected a friendly and informal style for daily activities, and a formal, professional, and neutral style for medical interactions \cite{tsiourti_virtual_2014}. Further, while older adults valued voice interactions \cite{santini_user_2021, jin_exploring_2024}, multi-modal input that combines voice, text, image, and video was preferred \cite{ter_stal_embodied_2020, jin_exploring_2024, bickmore_its_2005}. This preference is due to that voice interaction alone can be easily disrupted by other tasks, such as listening to music \cite{jin_exploring_2024}, or may not be suitable for screen-based activities, such as playing games \cite{stara_usability_2021}. 

\subsubsection{Appearance} \label{subsubsec:expect of appearance}
Five studies discussed older adults' expectations of CA appearance. Three of them noticed older adults' different perceptions of the gender appearance of CAs: Two studies found that the majority of their participants preferred female CAs in medication instruction \cite{azevedo_using_2018} and healthy diet coaching \cite{justo_analysis_2020}; One study showed that a male CA was perceived more authoritative than a female one in older adults' health status assessment \cite{ter_stal_embodied_2020}. The other two studies reported cultural differences in preference of CA appearances: \citet{tsiourti_virtual_2014} asked older adults' opinions about humanlike CA mockups and cartoonlike ones. They found that older adults from the Netherlands preferred realistic humanlike CAs; while those from Switzerland preferred fictitious cartoonlike CAs because they thought humanlike CAs are too impersonal and serious; Through semi-structured interviews, \citet{moller_user_2022} found that participants from Germany and Japan didn't expect CAs that look "too realistic" in order not to be fooled. 

\subsubsection{Personalization} \label{subsubsec: expect personalization}
Four studies identified that older adults have more personalized needs for CAs, including CA's appearance (e.g., clothing, gender, and age) \cite{santini_user_2021, tsiourti_virtual_2014}, information service \cite{santini_user_2021, brewer_empirical_2022, nallam_question_2020}, and communication style \cite{santini_user_2021, desai_painless_2023}. Regarding appearance some participants voiced that they should be able to tailor CA appearance by themselves \cite{santini_user_2021, tsiourti_virtual_2014}. For information services, \citet{nallam_question_2020} and \citet{brewer_empirical_2022} reported older adults' need for personalized information and recommendations based on knowledge of their health conditions and preferred information sources. For communication style, \citet{desai_painless_2023} focused on VUI and found older adults' need for personalization on the voice tone, pacing, and accent. 

\subsubsection{Sense of control} 
Four studies discussed older adults' desire for proactive versus reactive control over CAs. For example, \cite{santini_user_2021} found that older adults preferred reactive CAs rather than proactive ones, as they wanted to maintain full control over decisions such as not to engage in certain activities. \citet{tsiourti_virtual_2014} found that although some older adults preferred proactive functions, e.g., exercise reminder, they hoped that these functions could be completely deactivated by themselves. Older adults also wanted to control other activities of CAs. For example, they preferred the ability to turn off CAs as needed \cite{vardoulakis_designing_2012}. They also wanted to review and approve any health and well-being data before it was shared with doctors and hoped that CAs would present this information in a way that reflects a positive personality, e.g., "I'd like CAs to tell whoever that I'm a happy person." \cite{brewer_if_2022} 

\section{Discussion} \label{sec:discussion}
\subsection{Revisit Older Adult-CA Relationship: Advocating for Social Factors}
\subsubsection{Destigmatize CAs among Older Adults}
Stigmatization is a significant factor affecting older adults’ acceptance of conversational agents (CAs). Some older adults perceive the assistive features of CAs as markers of "being old and weak," which may deter adoption (see Section \ref{subsubsec:acceptance of CAs} and \ref{subsubsec:concerns of CA}). Similarly, assistive products like walkers or hearing aids can highlight older adults' physical or cognitive limitations, potentially stigmatizing older adults and undermining their sense of autonomy \cite{czaja_designing_2019, astell_thats_2020}. In contrast, products that emphasize older adults’ abilities are often more appealing. For instance, \citet{caldeira_i_2022} noted that devices like iPads may symbolize "the ability to learn new things" and are valued as representations of cognitive competence in later life.

To improve CA design, we propose reframing the focus from creating tools for "someone who needs help" to designing for "someone who is capable." This involves avoiding functionalities that merely substitute for users' activities and instead prioritizing features that empower older adults in their daily routines, enhancing their sense of competence and capability. To achieve this, we suggest the following design strategies:

(1) \textit{Promote lifelong learning for older adults}. For instance, in prior studies, \citet{desai_painless_2023} developed CAs to support health education for older adults, while \citet{bickmore_its_2005} introduced a CA that acts as a virtual exercise trainer.

(2) \textit{Enable interdependent social support}. CAs can play a pivotal role in fostering interdependent social connections by emphasizing social resources and encouraging meaningful engagement in community and family activities. By supporting older adults in managing and maintaining their social interactions, CAs can bolster emotional resilience and self-determination. For example, voice assistants and text-based chatbots can provide regular reminders about social events, family gatherings, and community activities. They can also facilitate consistent communication by prompting older adults to reach out to family members or friends and by simplifying access to tools like video calls or messaging platforms. Importantly, instead of seeking to replace human relationships, CAs act as facilitators that empower older adults to strengthen and sustain their social networks, thereby reducing isolation and reinforcing emotional well-being.

\subsubsection{Build Independent Older Adult-CA Relationship}
Over-reliance on CAs is a significant concern among older adults. However, many also acknowledge the convenience of relying on CAs to mitigate the unavoidable decline in physical and cognitive abilities associated with aging (see Section \ref{subsubsec:concerns of CA}). \citet{czaja_designing_2019} found that older adults are often willing to trade some degree of control and privacy for increased utility, particularly when such trade-offs enable them to maintain independent living. Importantly, older adults must be adequately informed to make these decisions. To better understand the relationship between aging and the level of assistance required, \citet{caldeira_i_2022} proposed a continuum of three stages of aging: 
(1) \textit{Active Coping}. During this stage, older adults can manage their daily activities with minimal assistance and may benefit from general-purpose technologies, such as smartwatches, primarily for safety monitoring; 
(2) \textit{Adaptive Coping}. This stage involves addressing specific impairments, such as mitigating the risks of falling, which necessitate targeted assistance technologies; 
(3) \textit{Passive Coping}. In this stage, older adults experience significant loss of autonomy and require extensive care and support. \citet{caldeira_i_2022} emphasized that technologies and services should be tailored to align with these stages of aging.

In the context of CAs, future research may explore ways to adapt CA functionality in accordance with these stages of aging. To illustrate this concept, we draw on the AI role taxonomy proposed by \citet{cimolino_two_2022}, which identifies five roles for AI in human-AI collaboration: 
(1) \textit{Supportive AI}: Assists users with specific tasks while keeping them in primary control. 
(2) \textit{Delegated AI}: Takes over specific tasks from the user. 
(3) \textit{Cooptable AI}: Performs certain tasks but remains under the user’s control. 
(4) \textit{Reciprocal AI}: Supports a dynamic relationship where tasks are delegated and re-assigned between the user and the AI.
(5) \textit{Complementary AI}: Fully handles specific portions of a task, leaving others to the user. 
Mapping these roles to the stages of aging, we suggest the following adaptation strategies for CAs:

\begin{itemize}
\item Active Coping: CAs should primarily act as supportive or cooptable agents, enabling older adults to maintain control over their tasks while receiving minimal, targeted assistance.
\item Adaptive Coping: CAs should facilitate task delegation, such as monitoring for symptoms of medical emergencies and initiating appropriate actions, such as contacting caregivers or emergency services.
\item Passive Coping: CAs should take on complementary roles, providing comprehensive support as part of extensive care for older adults.
\end{itemize}

\subsection{Develop CAs from Older Adults' Perspective}
\subsubsection{Optimize Older Adult-CA Interaction}
The fluency of interactions between older adults and conversational agents (CAs) significantly influences their acceptance of these technologies (see Section \ref{subsubsec:acceptance of CAs}). Additionally, we observed that older adults expressed concerns about the reliability of offline interactions (see Section \ref{subsubsec:concerns of CA}) and emphasized the importance of natural, intuitive communication with CAs (see Section \ref{subsubsec:expect natural conversations}). With the recent advancements in multi-modal large models, integrating these capabilities into future CAs could better meet older adults' needs for more natural, sustained, and long-term conversations. To further enhance Older Adult-CA interaction, we propose the following directions:
\begin{itemize}
    \item \textit{Offer instructional tutorials} to help older adults learn how to effectively interact with CAs.
    \item \textit{Enhance CAs’ ability to handle dialect-based conversations} to improve accessibility and understanding.
    \item \textit{Develop context-sensitive CAs} that can adjust their voice, terminology, and conversational style to align with specific contexts. Given older adults’ nuanced preferences for formal and informal conversational styles in different situations \cite{tsiourti_virtual_2014}, more detailed design guidelines are necessary.
    \item \textit{Enable multi-modal interaction options} to cater to varying preferences, such as voice-based interactions for gaming \cite{stara_usability_2021} or text-based interactions for music selection \cite{jin_exploring_2024}.
    \item \textit{Support offline interactions} to ensure functionality during events like Internet outages or power failures.
\end{itemize}

\subsubsection{Personalize CAs}
We identified older adults' requirements for personalized CAs in terms of appearance, information services, and communication styles (see Section \ref{subsubsec: expect personalization}). This aligns with findings from \citet{santini_user_2021}, where older adults emphasized that a \textit{one-size-fits-all} approach is unsuitable for CA design. They preferred CAs tailored to their unique needs across different contexts and activities. Considering the diversity within the aging population, we argue that personalizing various CA functions is essential. Such personalization enhances older adults' perception of usefulness, thereby improving their acceptance of CAs \cite{czaja_designing_2019}.

Based on prior research into CA personalization, we outline two primary approaches:
(1) Allowing older adults to define their own CAs. This includes customizing aspects like appearance, voice, and conversational style \cite{santini_user_2021, tsiourti_virtual_2014}. Such customization enhances ease of use and leads to greater acceptance, as highlighted by \citet{czaja_designing_2019}.
(2) Developing adaptability in CAs. This approach involves designing CAs that can adjust to older adults' aging status and preferences. It requires CAs to recognize the user’s functional, psychological, and social aging stages through interactions and tailor functionalities accordingly. To operationalize aging stages, researchers may consider dimensions beyond chronological age, including functional age, which reflects physical and functional, psychological age, which measures cognitive and emotional maturity, and social age, which accounts for roles and behaviors relative to societal norms \cite{settersten_measurement_1997, johfre_social_2023}.

However, addressing the diverse needs of older adults through personalization presents a potential dilemma. Developing adaptable CAs necessitates data collection and analysis to understand aging stages and preferences, which raises concerns about privacy — a major concern for older adults (see Section \ref{subsubsec:concerns of CA}). Possible solutions to this challenge are discussed in the following subsection.

\subsubsection{Tackle the Challenges of Privacy Protection}
Older adults emphasize the importance of privacy protection and the need for transparency in data recording and sharing (see Section \ref{subsubsec:concerns of CA}). 
%The emergence of LLM-powered CAs introduces new privacy challenges. Because LLMs are trained on data aggregated from multiple sources, including user interactions, there is a risk that they "inadvertently" disclose personal information during responses, as reported by \cite{white_how_2023}. 
Further, some older adults have expressed a preference for anthropomorphic CA appearances (see Section \ref{subsubsec:expect of appearance}) and humanized conversational styles (see Section \ref{subsubsec:expect natural conversations}). Research indicates that anthropomorphic CAs may elicit anthropomorphizing behaviors from users, which can result in privacy risks analogous to those found in personal relationships \cite{mooradian_conversational_2024}. In such cases, users may unconsciously share sensitive personal data without fully realizing or controlling the extent of their disclosure. To address these challenges, we propose the following directions for HCI scholars:

(1) \textit{Enhancing Transparency in Data Recording and Sharing}.
Transparent practices should include designing clear consent mechanisms for data collection, providing tutorials to teach older adults how to check their data status and manage data collection and sharing, and incorporating signals within CAs to indicate their data collection and sharing status. Designers should consider delivering these features through natural language conversations, as this approach may be more intuitive and accessible for older adults.

(2) \textit{Developing Tutorials for Confident CA Use}. 
Tutorials should aim to educate older adults on how to use CAs while maintaining their privacy. For instance, CAs could include contextual hints in their responses, such as providing reminders to avoid sharing sensitive information (e.g., "To generate a self-introduction while protecting your privacy, consider using placeholder names for institutions").

(3) \textit{Exercising Caution in Anthropomorphic CA Design}. 
Since research indicates that older adults frequently anthropomorphize CAs \cite{chin_like_2024} and may form trust similar to relationships with friends or family, they could unintentionally disclose personal information. It is essential to conduct further research to carefully balance the benefits and risks of anthropomorphic CA design, particularly in contexts where trust could lead to unintended privacy breaches.

\subsubsection{Empower CAs with multi-agent systems}
This review reveals that prior research has addressed a wide array of health domains and explored the diverse roles of CAs, such as serving as companions and coaches (see Section \ref{subsubsec:roles of CAs}). Additionally, studies have documented older adults’ expectations for various CA functionalities, including medication management and mental health support (see Section \ref{subsubsec:funtionalities}). These insights underscore numerous potential use cases for CAs and suggest that, to effectively meet older adults’ varied expectations, CAs should be equipped with: 
(1) Knowledge across diverse health domains, 
(2) Workflows tailored to specific health activities, and
(3) Communication capabilities to facilitate collaboration with relevant stakeholders (e.g., healthcare providers for data sharing and external assistance).

We propose multi-agent systems as a promising solution to address these complex requirements. Recent advancements have begun exploring the potential of multi-agent conversational systems powered by LLMs to meet users’ multifaceted needs. For instance, \citet{lippert_multiple_2020} introduced a multi-agent conversational system designed to support student learning by enabling interaction with multiple agents, each providing specialized knowledge. Similarly, future CAs designed for older adults’ health management could adopt a multi-agent architecture, where individual agents are responsible for addressing specific health conditions or needs. This approach could allow for a more modular and specialized system, better equipped to handle the diverse and dynamic health challenges older adults face.

\subsubsection{Safeguard a sustainable positive effect of CAs on older adults' health}
In this review, half of the included studies examined the impact of CAs on older adults’ health. Although the majority reported positive effects, these findings were not grounded in robust longitudinal evidence. Several studies indicated that the benefits of CAs might diminish over time, potentially due to declining novelty or environmental challenges (see Section \ref{subsec:effects of CAs}). This highlights a key challenge for current CA development: ensuring that the positive effects of CAs on older adults’ health are sustainable over the long term.

Furthermore, we observed a technological shift from traditional rule-based chatbots to voice assistants in earlier studies and, more recently, the integration of generative AI into CA systems (see Section \ref{subsubsec:types of CAs}). This progression in CA technology, combined with varying research methodologies, complicates the comparison and validation of user study results. Findings derived from older chatbot systems may not align with the outcomes of newer, AI-enhanced CAs. On one hand, advanced CAs may directly address some of the functional limitations of earlier systems. On the other hand, subjective factors such as trust, privacy concerns, and the risk of over-dependency may persist across both types of systems. Therefore, it is critical to identify and account for diverse factors, such as CA design variations, sample biases, intervention durations, measurement inconsistencies, and the influence of external events that contribute to the potentially mixed or conflicting effects observed in user studies.

We recommend the following approach to identify factors influencing the sustainability of CAs' effect on older adults health: 
(1) Develop a comprehensive evaluation framework to assess the performance and impact of CAs on older adults in healthcare. While some initial work has proposed metrics for evaluating and comparing CA performance \cite{venkatesh_evaluating_2017}, limited research has focused specifically on healthcare applications for older adults. %Additionally, although most studies report positive outcomes from CA use (see Section \ref{subsec:effects of CAs}), it remains unclear how enduring these effects are. Longitudinal studies have noted declines in CA utility over time \cite{bickmore_randomized_2013} and reduced user engagement among older adults \cite{lima_discovering_2023}. To address these gaps, we recommend 
(2) Adopt longitudinal research approaches to evaluate the sustained effects of CAs on older adults’ health outcomes, enabling the design of more effective and enduring solutions.
%In almost 90\% percent of the included studies, more than half of participants are female (see Section \ref{}). gender, race, health condition, educational background and other factors (see Section \ref{subsec:participants reviewed}). We found that more studies are needed to investigate older adults who are male, have lower levels of education, and lower incomes, to better address the coverage of the aging population. 

\section{Conclusion} \label{sec:conclusion}
This paper presents a systematic review of 72 studies on conversational agent (CA) research for older adults’ health. It highlights key research trends, including the transition from embodied conversational agents (ECAs) to voice assistants and chatbots, as well as the diverse roles that CAs play in supporting older adults’ health, such as acting as coaches, companions, and general health assistants. The review identifies several challenges for future research. These include the mixed and sometimes conflicting effects of CAs on older adults, their relatively low acceptance and concerns regarding CAs, and their expectations for CAs to offer multi-functional support, engage in natural conversations, enable personalization, and preserve their sense of control.
To address these challenges, the paper advocates rethinking the relationship between older adults and CAs by reducing stigmatization and fostering independent interactions between them. Additionally, it recommends designing CAs from the perspective of older adults, with particular emphasis on optimizing interaction design, enhancing personalization, protecting privacy, expanding functionality, and ensuring positive outcomes. These recommendations aim to inform future research and development efforts in creating CAs that effectively address the healthcare needs of older adults.

\bibliographystyle{ACM-Reference-Format}
\bibliography{references_group, additional, ChatGPT_references}

\appendix

\section{Literature Search Queries} \label{appendix:search queries}

\paragraph{Query variations searched for CAs} “conversational agent”, “conversational AI”, ‘‘conversational partner’’OR “conversational system”, “conversatonal-agent based system”, “dialog system”, “dialogue system”, “coaching system”, "chatbot", "chatterbot", "chatterbox", “communication robot”, “communicative machine”, “intelligent assistant”, “voice assistant”, “digital assistant”, “virtual assistant”, “robotic assistant”, “virtual agent”, ”intelligent agent”, “relational agent”, “digital agent”, “social agent”, “companion agent”, “screen agent”, “smart speaker”, “coach*”, “e-coach*”, “robot*”, “virtual companion”, “virtual carer”, “virtual educator”, “virtual expert”, “virtual friend”, “virtual instructor”, “virtual mentor”, “virtual personal trainer”, “virtual therapist”, “virtual tutor”, “digital pet”, “robotic pet”, “robotic psychological assistance”, “socially communicative machine”, “home dialogue system” and “digital avatar”

\paragraph{Query variations searched for health} “health”, “healthcare”, “wellbeing”, “fitness”, “well-being”, “eHealth”, “telehealth”, “telecare”, “mHealth” and “healthy aging”

\paragraph{Query variations searched for older adults} “older adult”, “elder*”, “aging”, “grownup”, “senior”, and “retiree”

\section{Publication count by year and venue} \label{appendix:publication count}

\begin{landscape}
\begin{table}. 
    \caption{Publication count by year and venue}
    \setlength{\arrayrulewidth}{0mm}
    \centering
    \begin{tabular}{p{5cm}|llllllllllr}
    \toprule
    Venue & 2005 - 2015 & 2016 & 2017 & 2018 & 2019 & 2020 & 2021 & 2022 & 2023 & 2024 & Total \\
    \midrule
    Conference on Human Factors in Computing Systems (CHI) &  &  &  &  &  &  & 2 & 3 & 1 & 1 & 7 \\
    AMIA Annual Symposium Proceedings (AMIA) &  & 1 &  & 1 &  &  &  &  &  &  & 2 \\
    Gerontology and Geriatric Medicine &  &  &  &  &  & 1 &  &  & 1 &  & 2 \\
    International Journal of Social Robotics &  &  &  &  &  &  &  & 2 &  &  & 2 \\
    International journal of environmental research and public health &  &  &  &  &  &  & 1 & 1 &  &  & 2 \\
    JMIR Aging &  &  &  &  &  &  &  & 1 &  & 1 & 2 \\
    JMIR formative research &  &  &  &  &  &  &  & 1 &  & 1 & 2 \\
    Journal of Medical Internet Research (JMIR) &  &  &  &  & 1 & 1 &  &  &  &  & 2 \\
    Proceedings of the ACM on Human-Computer Interaction &  &  &  &  &  & 1 &  & 1 &  &  & 2 \\
    Others & 12 & 1 & 2 & 1 & 2 & 6 & 7 & 8 & 7 & 2 & 48 \\
    \midrule
    Total & 12 & 2 & 2 & 2 & 3 & 9 & 10 & 18 & 9 & 5 & 72 \\
    \bottomrule
    \end{tabular}
    \label{tab:pub_count}
\end{table}
\end{landscape}

\section{Participant characteristics in included studies} \label{appendix:p_characteristics}
\begin{table} [ht]
    \caption{Summary of participant characteristics}
    \setlength{\arrayrulewidth}{0mm}
    \centering
    \begin{tabular}{@{}l|l|l}
    \toprule
        \textbf{Characteristics} & \textbf{Number of studies} & \textbf{\%})\\
    \midrule
        \textbf{Sample size} & & \\
          1-50 & 65 & 83.33\\
          50-100 & 8 & 10.26\\
          >100 & 5 & 6.41\\
        \textbf{Minimum age of participants} &  & \\
          50-59 & 9 & 11.54\\
          60-69 & 29 & 31.18\\
          70-79 & 7 & 8.97\\
          Unreported & 33 & 42.31\\
        \textbf{Maximum age of participants} &  & \\
          60-69 & 1 & 1.28 \\
          70-79 & 11 & 14.10\\
          80-89 & 19 & 24.36\\
          90-99 & 14 & 17.95\\
          Unreported & 33 & 42.31\\
        \textbf{Average age of participants} &  & \\
          50-59 & 1 & 1.28\\
          60-69 & 12 & 15.38 \\
          70-79 & 25 & 32.05\\
          80-89 & 3 & 3.85\\
          Unreported & 37 & 47.44\\
        \textbf{Percentage of female participants} &  & \\
          <20\% & 1 & 1.28\\
          20\%-40\% & 2 & 2.56\\
          40\%-60\% & 21 & 26.92\\
          60\%-80\% & 26 & 33.33\\
          80\%-100\% & 12 & 15.38\\
          100\% & 4 & 5.13\\
          Unreported & 12 & 15.38\\
       \textbf{Percentage of participants with high school and lower education} &  & \\
          <20\% & 7 & 8.97\\
          20\%-40\% & 6 & 7.69\\
          40\%-60\% & 6 & 7.69\\
          60\%-80\% & 4 & 5.13\\
          80\%-100\% & 2 & 2.56\\
          100\% & 2 & 2.56\\
          Unreported & 51 & 65.38\\
       \textbf{Race} &  & \\
          Multi-races & 18 & 23.08\\
          Asian & 6 & 7.69\\
          Black or African American & 1 & 1.28\\
          White or Caucasian & 1 & 1.28\\
          Unreported & 52 & 66.66\\
       \textbf{Health condition} &  & \\
          With health conditions & 14 & 17.95\\
          Healthy in specific aspects & 9 & 11.54\\
          With different levels of health status & 8 & 10.26\\
          Healthy & 7 & 8.97\\
          Unreported & 40 & 51.28\\
       \textbf{Living condition} &  & \\
          Live alone & 14 & 17.95\\
          Multiple living status & 9 & 11.54 \\
          Live in care facilities & 7 & 8.97 \\
          Live with caregivers & 2 & 2.56 \\
          Live in communities & 1 & 1.28 \\
          Live in a single-family home & 1 & 1.28\\
          Unreported & 44 & 56.41 \\
    \bottomrule
    \end{tabular}
    \label{tab:p_characteristics}
\end{table}

\end{document}